\documentclass[aps,twocolumn,showpacs,amssymb,amsmath]{revtex4}
\usepackage{graphicx}
\usepackage{color}

\begin{document}

\title{Mode-mode coupling theory of resonant pumping via dynamical tunneling processes in a deformed microcavity}
\author{Kyungwon An}
\email{kwan@phya.snu.ac.kr}
\author{Juhee Yang}
\affiliation{School of Physics and Astronomy, Seoul National University, Seoul 151-742, Korea}
\date{\today}
\begin{abstract}
Mode-mode coupling theory is presented for the resonant pumping via dynamical tunneling processes in a deformed microcavity. From the steady-state solution of the coupled differential equations of uncoupled chaotic modes and an uncoupled high-Q regular mode, pumping efficiency is obtained as a function of pump detuning, coupling constants and decay rates of the involved uncoupled modes. We show that the pump-excited chaotic modes as a whole can be regarded as a single pump mode with an effective decay rate and an effective coupling constant with the regular mode. We also show that the decay rate of the regular mode is enhanced by dynamical tunneling into all chaotic modes. Analysis method to obtain the effective coupling constant from the pumping efficiencies is presented for a two-dimensional deformed microcavity.
\end{abstract}


\maketitle

\section{Introduction}
It is well known that a high-Q mode in a deformed microcavity can decay by chaos-assisted dynamical tunneling. High-Q modes are usually localized in regular regions in a phase space surrounded by chaotic sea. Without chaos-assisted dynamical tunneling the only way that the light in a high-Q mode can escape the cavity is by directly tunneling to the outside continuum. The output resulting from this is usually very weak. A strong directional output can be obtained if there exists a chaos-assisted dynamical tunneling process in action: the light in a high-Q mode can undergo dynamical tunneling into nearby chaotic sea first and then go through chaotic ray dynamics until it escapes the cavity by ray refraction at particular positions and at particular angles. 

Quite recently, Yang {\em et al.}\,\cite{resonant-pumping} has reported a resonant pumping experiment based on dynamical tunneling from chaotic sea to high-Q modes in a deformed microcavity laser. They introduced a collimated pump beam into the cavity by refraction in a time reversed way with respect to the directional output caused by dynamical tunneling followed by chaotic ray dynamics. They observed that the pumping efficiency of a high-Q lasing mode at a longer wavelength than that of the pump is enhanced by two orders of magnitude whenever the pump is resonant with a high-Q cavity mode, which is localized in a regular region in the phase space separated from the chaotic sea. Since the pump beam, injected by refraction, moves in the chaotic sea, the resonant enhancement must have come from the dynamical tunneling from the chaotic sea to the regular mode.

In this paper, we present a mode-mode coupling theory for this pumping process. We first set up coupled differential equations between uncoupled chaotic modes and an uncoupled high-Q mode with the chaotic modes driven by a pump beam in Sec.\,\ref{sec2}. From the steady-state solution of the coupled differential equations, we derive a formula for relative pumping efficiency as a function of pump detuning, coupling constants and decay rates of the involved modes in Sec.\,\ref{sec3}. We show that the pump-excited chaotic modes can be regarded as a single pump mode with an effective decay rate and an effective coupling constant $\bar g$ with the regular mode in Sec.\,\ref{sec3C}. We also show that the decay rate of the regular mode is enhanced by dynamical tunneling into all chaotic modes. The same result is obtained by considering the eigenvalue problem associated with the coupled differential equations in Sec.\,\ref{sec4}. We finally describe analysis method to obtain the effective coupling constant $\bar g$ from the experimentally measured pumping efficiencies and apply it to the recent work by Yang {\em et al.}\,\cite{resonant-pumping} as an example in Sec.\,\ref{sec5}.

\section{Mode-mode coupling model} \label{sec2}
\begin{figure}
\includegraphics[width=3.4in]{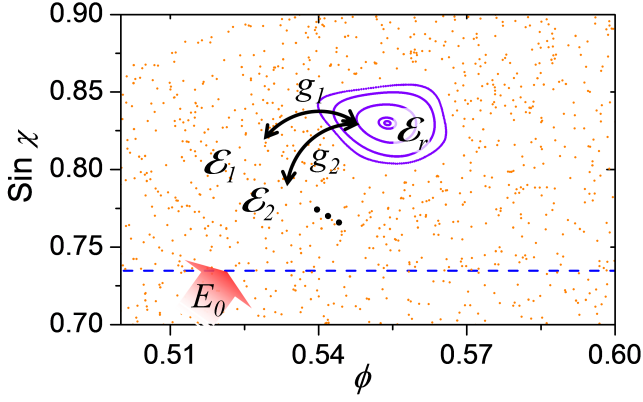}
\caption{Uncoupled chaotic modes, excited by refractively injecting a beam of a pump laser, can couple to an uncoupled regular mode by dynamical tunneling processes. The $n$th uncoupled low-Q chaotic mode of an envelope $\mathcal{E}_n$ can be coupled to an uncoupled high-Q regular mode of an envelope $\mathcal{E}_r$ with a coupling constant $g_n$ ($n=1,2,3,\ldots,N$). The amplitude of a driving field is $E_0$ with a coupling coefficient $a_n$ to the $n$th chaotic mode.} 
\label{fig1}
\end{figure}

Dynamical tunneling between an {\em uncoupled} regular mode and {\em uncoupled} chaotic modes can be modeled as mode-mode coupling, as depicted in Fig.\,\ref{fig1}. Here, the uncoupled modes (or states if you will) are not true eigenmodes of the system since the coupling between them are treated separately in our model -- the term ``mode'' from now on actually means an uncoupled mode/state if not noted otherwise. 
Many chaotic modes (low-Q cavity modes) of decay rates $\gamma_n$ ($n=1,2,3,\cdots,N$) are driven by a pump laser and these modes can then be coupled to a regular mode (a high-Q cavity mode) of a decay rate $\gamma_r$ by dynamical tunneling processes.

Let us write the electric field of the $n$th chaotic mode as $E_n(\mathbf x, t)=\mathcal E_n(t)f_n(\mathbf x) e^{-i\omega t}$ and that of the regular mode as $E_r(\mathbf x, t)=\mathcal E_r(t)f_r(\mathbf x) e^{-i\omega t}$. Here $\mathcal E_n$ and $\mathcal E_r$ are their slowly-varying envelopes in time and $f_n$ and $f_r$ are their normalized spatial mode functions, respectively. 
We assume that the chaotic modes have been obtained by a proper unitary transformation among them in such a way that they are orthogonal to each other. Hence, there exists no coupling between any two of them.
\begin{displaymath}
\int f_p^*(\mathbf x) f_q (\mathbf x)d^3x=\delta_{pq}\quad {\rm where\;} p,q=1,2,\ldots,N
\end{displaymath}
The equations of motion for envelopes $\mathcal E_n$ and $\mathcal E_r$ are basically those of driven coupled oscillators \cite{semiclassical}:
\begin{eqnarray}
\dot{\mathcal E}_n +
\gamma_n\mathcal{E}_n&=&a_nE_0-g_n\mathcal{E}_r, \label{eq1}
\\
\dot{\mathcal E}_r + (\gamma_r+i\Delta)
\mathcal{E}_r&=&\sum^N_ng_n\mathcal{E}_n, \label{eq2}
\end{eqnarray}
where $g_n$ is the coupling constant (assumed real) between the $n$th chaotic mode and the regular mode.
We assume that the chaotic modes are extremely lossy while the coupling of each to the regular mode is so weak that the oscillation between any chaotic mode and the regular mode is over-damped: $|g_n| \ll \gamma_n$. 
Effects of partial barriers \cite{Backer09} are included in $g_n$ and $\gamma_n$.
Symbol $a_n$ is the coupling coefficient of the external pump laser into the $n$th chaotic mode, depending on the position and the incident angle of the pump beam on the cavity boundary, and $\Delta \equiv \omega-\omega_r$, the detuning between the pump laser  of frequency $\omega$ and the regular mode of frequency $\omega_r$. Since each chaotic mode is so broad in linewidth and since we are interested in a small frequency range ($\sim\gamma_r$) around the regular mode, we have neglected a detuning term for the chaotic mode in writing Eq.\,\eqref{eq1}.

\section{Steady state solution} \label{sec3}

By letting all $g_n=0$, we simulate the non-resonant pumping. In this case in the steady state, we get
\begin{equation}
\mathcal E_n=\frac{a_n}{\gamma_n}E_0\equiv \mathcal E_n^0,\quad
E_p(\mathbf x)=\sum_n\mathcal E_n^0 f_n(\mathbf x) = \mathcal E_p^0 f_p(\mathbf x),
\end{equation}
where
\begin{equation}
f_p(\mathbf x)\equiv\frac{1}{\mathcal E_p^0}\sum_n\mathcal E_n^0 f_n(\mathbf x)\quad {\rm with}\quad \mathcal E_p^0\equiv \sqrt{\sum_n |\mathcal E_n^0|^2}
\label{eqfp}
\end{equation} 
such that
\begin{displaymath}
\int f_p^*(\mathbf x)f_p(\mathbf x)d^3\mathbf x=1.
\end{displaymath}
The quantity $\mathcal E_n^0$ is the steady-state amplitude of the $n$th chaotic mode excited by the pump laser of $E_0$. Only a small portion of the chaotic modes that have appreciable $a_n/\gamma_n$ among a large number of chaotic modes make up the function $f_p(\mathbf x)$. Therefore, $f_p(\mathbf x)$ can be regarded as the (normalized) mode function of the nonresonant intracavity pump field or a pump mode in short. 

In the experiment of Yang {\em et al.}\,\cite{resonant-pumping}, the position and the incident angle of the focused beam of the pump laser were adjusted in order to maximize the pumping efficiency. This procedure is nothing but optimizing the set $\{a_n \}$ or the pump mode for the strongest tunneling to the regular mode.

For nonzero $g_n$, the steady-state solution is
\begin{eqnarray}
\mathcal E_r &=& \frac{\bar{g}}{\gamma_r+i \Delta}\frac{\mathcal
E_p^0}{1+\mathcal{G}}, \label{eq3}
\\
\mathcal E_n&=&\mathcal E_n^0 - \mathcal G_n \frac{\mathcal E_p^0}{1+\mathcal G}
=\frac{\mathcal E_n^0+\left[\mathcal E_n^0 \mathcal G-\mathcal E_p^0 \mathcal G_n\right]}{1+\mathcal G}\simeq \frac{\mathcal E_n^0}{1+\mathcal G},\nonumber\\
\label{eq4}
\end{eqnarray}
where
\begin{eqnarray}
\bar g &\equiv& \frac{1}{\mathcal E_p^0} \sum_n g_n \mathcal E_n^0, \label{eq6}
\\
\mathcal G &\equiv& \frac{1}{\gamma_r+i\Delta}\sum_ng_n^2/\gamma_n, 
\\
\mathcal G_n &\equiv& \frac{g_n \bar{g}}{\gamma_n(\gamma_r+i\Delta)}.
\end{eqnarray}
The quantity in [\dots] was neglected in Eq.\,\eqref{eq4} since
\begin{eqnarray}
\mathcal E_n^0 \mathcal G-\mathcal E_p^0 \mathcal G_n
=\frac{1}{\gamma_r+i\Delta}\sum_m \left(\mathcal E_n^0 \frac{g_m}{\gamma_m}-\mathcal E_m^0 \frac{g_n}{\gamma_n}\right)g_m,
\end{eqnarray}
the magnitude of which is much smaller than $|\mathcal E_n^0 \mathcal G|$, and since $|\mathcal G|$ is at most of order of unity as to be seen later, the quantity in the square brackets can be safely neglected. The pump mode is then obtained from
\begin{eqnarray}
E_p(\mathbf x)=\sum_n\mathcal E_n f_n(\mathbf x) = \mathcal E_p f_p(\mathbf x)\quad{\rm with}\quad
\mathcal E_p\equiv\frac{\mathcal E_p^0}{1+\mathcal G}, \nonumber\\
\label{Ep0}
\end{eqnarray}
where $f_p(\mathbf x)$ is defined by Eq.\,\eqref{eqfp}.

We are interested in the intensities, $I_r\equiv|\mathcal E_r|^2$ and $I_p\equiv|\mathcal E_p|^2$, for the evaluation of which we introduce the following quantities:
\begin{equation}
\frac{1}{\gamma_p}\equiv \sum_n\frac{1}{\gamma_n}\left(\frac{g_n}{\bar g}\right)^2, \label{eq10}
\end{equation}
and
\begin{equation}
G\equiv\sum_n\frac{g_n^2}{\gamma_n\gamma_r}. \label{eq11}
\end{equation}
We then obtain the intensities of the regular mode and the pump mode, respectively, as
\begin{eqnarray}
I_r(\delta)&=&I_p^0\frac{\gamma_p}{\gamma_r}\frac{G}{(1+G)^2+\delta^2}, \label{eq12} \\ 
I_p(\delta)&=&I_p^0\frac{1+\delta^2}{(1+G)^2+\delta^2}, \nonumber\\
&=& I_p^0 \left[1-\frac{G(2+G)}{\delta^2+(1+G)^2}\right], \label{eq13}
\end{eqnarray}
where
\begin{equation}
I_p^0 \equiv |\mathcal E_p^0|^2, \quad \delta=\Delta/\gamma_r\;.
\end{equation}

The relative pumping efficiency $\epsilon$ with respect to the non-resonant pumping efficiency can be written as
\begin{equation}
\epsilon(\delta)=\frac{\int  \left[I_p(\delta)|f_p(\mathbf x)|^2+I_r(\delta)|f_r(\mathbf x)|^2\right]|f_l(\mathbf x)|^2 d^3\mathbf x}{\int I_p^0|f_p(\mathbf x)|^2|f_l(\mathbf x)|^2 d^3\mathbf x},  \label{eq15}
\end{equation}
where $f_l(\mathbf x)$ is the mode function of the lasing mode at which the pumping efficiency is measured in the work of Yang {\em et al.}\,\cite{resonant-pumping}. We introduce mode-overlap factors $\beta_p$ and $\beta_r$ defined as
\begin{equation}
\beta_p\equiv \int |f_p(\mathbf x) f_l(\mathbf x)|^2 d^3\mathbf x, \quad 
\beta_r\equiv \int |f_r(\mathbf x) f_l(\mathbf x)|^2 d^3\mathbf x\;.
\end{equation}
Using Eqs.\,\eqref{eq12} and \eqref{eq13}, we then  obtain
\begin{equation}
\epsilon(\delta)=\left[1-\alpha\mathcal L(\delta)\right]+\frac{\gamma_p\beta_r}{\gamma'_r\beta_p}\alpha'\mathcal L(\delta), \label{eq16}
\end{equation}
where a modified decay rate $\gamma'_r$, coupling efficiencies $\alpha$, $\alpha'$ and a unity-peak-normalized lineshape $\mathcal L (\delta)$ are defined as
\begin{eqnarray}
\gamma'_r&\equiv& \gamma_r(1+G), \label{eq17}
\\
\alpha&\equiv&\frac{G(2+G)}{(1+G)^2},
\\
\alpha'&\equiv&\frac{G}{(1+G)},
\\
\mathcal L (\delta)&=&\frac{(1+G)^2}{\delta^2 +(1+G)^2}\;. \label{eq20}
\end{eqnarray}
The coupling efficiency $\alpha$ is defined as the fraction of the intracavity pump power of non-resonant pumping that is reduced in the case of the resonant pumping, similarly defined to the case of tapered-fiber couplers \cite{tapered-fiber}. The other coupling efficiency $\alpha'$ is a fraction of the intracavity pump power of non-resonant pumping that is transferred to the regular mode in the resonant pumping. The physical meaning of $\gamma'_r$ is to be discussed in Sec.\,\ref{sec3C}.

\subsection{Effect of linewidth of the pump laser}

The above results were obtained for a monochromatic pump laser. If the pump has a linewidth $\gamma_L$ comparable to or larger than
the linewidth $\gamma_r$ of the high-Q cavity, we should average the results over the spectral lineshape of the pump 
\begin{equation}
\mathcal L_L(\omega')=\frac{\gamma_L^2}{(\omega'-\omega)^2+\gamma_L^2} \nonumber
\end{equation}
with $\omega$ interpreted as its center frequency. The lineshape function in Eq.\,\eqref{eq20} then becomes
\begin{equation}
\mathcal L(\delta) \rightarrow \pi (1+G)
\int \frac{(1+G)/\pi}{\delta'^2+(1+G)^2}\frac{(\gamma_L/\gamma_r)/\pi}{(\delta'-\delta)^2+(\gamma_L/\gamma_r)^2}d\delta',
\end{equation}
where $\delta'=(\omega'-\omega_r)/\gamma_r$ and $\delta=(\omega-\omega_r)/\gamma_r$. 

If $\gamma_L\ll \gamma_r$, the second integrand can be approximated as a delta function and thus we recover Eq.\,\eqref{eq20}. If $\gamma_L\gg \gamma'_r=\gamma_r(1+G)$ on the other hand, the first integrand is like a delta function.
\begin{eqnarray}
\mathcal
L(\delta)&\rightarrow&\pi(1+G)\frac{1}{\pi}\frac{(\gamma_L/\gamma_r)}{\delta^2+(\gamma_L/\gamma_r)^2}
\nonumber\\
&=&\left(\frac{\gamma'_r}{\gamma_L}\right)\mathcal
L_L(\delta)
\end{eqnarray}
The linewidth of the resonance appears to be that of the broad pump laser and there is also a reduction factor $\gamma'_r/\gamma_L\ll1$. Eq.\,\eqref{eq16} is now replaced in this limit with
\begin{equation}
\epsilon(\delta)=\left[1-\left(\frac{\gamma'_r}{\gamma_L}\right)\alpha \mathcal L_L(\delta)\right]+\frac{\gamma_p\beta_r}{\gamma_L\beta_p}\alpha'\mathcal L_L(\delta).
\label{eq26}
\end{equation}

\subsection{Multiple interference picture}
The steady-state solution given by Eqs.\,\eqref{eq3} and \eqref{eq4} can also be obtained from consideration of multiple interference of coupled fields. 
Without the coupling, $g_n=0$, the chaotic modes are excited to $\mathcal{E}_n=\mathcal{E}_n^0$ with no excitation of the regular mode, $\mathcal{E}_r=0$. In the presence of the coupling, $g_n\ne0$, the chaotic modes can tunnel to the regular mode. Let us consider the zero detuning case, $\Delta=0$, for simplicity. Equation \eqref{eq2} dictates that the amplitude of the regular mode tunneled from the $n$th chaotic mode is
\begin{displaymath}
\frac{g_n}{\gamma_r}\mathcal{E}_n^0
\end{displaymath}
with $g_n/\gamma_n$ as a tunneling coefficient composed of the coupling constant $g_n$ and the decay rate of the {\em destination} mode $\gamma_r$. Summing up the contributions from all chaotic modes, we obtain the first round contribution to the regular mode:
\begin{equation}
\mathcal{E}_r^{(1)}=\sum_n \frac{g_n}{\gamma_r}\mathcal{E}_n^0. \label{eq-Er1}
\end{equation}

This amplitude can tunnel back to the $m$th chaotic mode, the amplitude of which is given by
\begin{displaymath}
\mathcal{E}_m^{(1)}=\left(-\frac{g_m}{\gamma_m}\right)\mathcal{E}_r^{(1)},
\end{displaymath}
where the tunneling coefficient $(-g_m/\gamma_m)$ is dictated by Eq.\,\eqref{eq1}. Note the decay rate in the denominator is $\gamma_m$ since the destination this time is the $m$th chaotic mode. A fraction of this amplitude given by
\begin{displaymath}
\left(\frac{g_m}{\gamma_r}\right)\mathcal{E}_m^{(1)}=-\frac{g_m^2}{\gamma_m\gamma_r}\mathcal{E}_r^{(1)},
\end{displaymath}
can tunnel back to the regular mode. Summing up the contributions from all chaotic modes, we obtain the second round contribution:
\begin{displaymath}
\mathcal{E}_r^{(2)}=-\sum_m \frac{g_m^2}{\gamma_r\gamma_m}\mathcal{E}_r^{(1)}=-G\mathcal{E}_r^{(1)}.
\end{displaymath}
Extending this argument, we easily find the $k$th round contribution to be
\begin{eqnarray}
\mathcal{E}_r^{(k)}&=&-G\mathcal{E}_r^{(k-1)}, \\ \nonumber
\mathcal{E}_n^{(k)}&=-&\frac{g_n}{\gamma_n}\mathcal{E}_r^{(k)}. \nonumber
\end{eqnarray}

Summing up all multiple round contributions, we then obtain the amplitude of the regular mode in the form of multiple interference.
\begin{eqnarray}
\mathcal{E}_r&=&\mathcal{E}_r^{(1)}+\mathcal{E}_r^{(2)}+\mathcal{E}_r^{(3)}+\cdots \nonumber \\
&=&\mathcal{E}_r^{(1)}(1-G+G^2-\cdots)=\frac{\mathcal{E}_r^{(1)}}{1+G}=\frac{\sum_n g_n \mathcal{E}_n^0/\gamma_r}{1+G}, \nonumber
\end{eqnarray}
recovering Eq.\,\eqref{eq3} with $\Delta=0$. Likewise, the amplitude of $n$th chaotic mode is obtained by summing up all multiple round contributions in the form of multiple interference:
\begin{eqnarray}
\mathcal{E}_n&=&\mathcal{E}_n^0+\mathcal{E}_n^{(1)}+\mathcal{E}_n^{(2)}+\cdots \nonumber \\
&=&\mathcal{E}_n^{0}-\frac{g_n}{\gamma_n}\left(\mathcal{E}_r^{(1)}+\mathcal{E}_r^{(2)}+\mathcal{E}_r^{(3)}+\cdots\right) \nonumber\\
&=&\mathcal{E}_n^{0}-\frac{g_n}{\gamma_n} \mathcal{E}_r,\nonumber
\end{eqnarray}
which is nothing but Eq.\,\eqref{eq4} with $\Delta=0$. 

\subsection{Physical meaning of $\bar{g}$, $\gamma_p$ , $G$ and $\gamma'_r$} \label{sec3C}

In Eqs.\,\eqref{eq6} and \eqref{eq10}, $\bar{g}$ and $\gamma_p$ were defined, respectively, without identification of their physical meanings. We show below that $\bar{g}$ is the coupling constant between the regular mode of $\mathcal{E}_r$ and the pump mode of $\mathcal{E}_p$ 
and that $\gamma_p$ is the decay rate of the pump mode.

Consider the first round contribution to the regular mode in Eq.\,\eqref{eq-Er1}. By using the definition of $\bar g$ of Eq.\,\eqref{eq6}, this can be rewritten as
\begin{equation}
\mathcal{E}_r^{(1)}=\frac{\bar g}{\gamma_r}\mathcal{E}_p^0 \label{eq-Er1'}
\end{equation}
indicating that it can be regarded as a result of tunneling from the pump mode of $\mathcal{E}_p^0$ with a coupling constant $\bar g$.

We can compose the first round correction in the pump mode from the first round correction in the $n$th chaotic mode. From Eq.\,\eqref{eq4} with $\Delta=0$
\begin{displaymath}
\mathcal{E}_n\simeq \frac{\mathcal{E}_n^0}{1+G}\simeq \mathcal{E}_n^0-G\mathcal{E}_n^0+G^2\mathcal{E}_n^0-\cdots
\end{displaymath}
and thus 
\begin{displaymath}
\mathcal{E}_n^{(1)}\simeq -G\mathcal{E}_n^0\;.
\end{displaymath}
The first round correction in the pump mode is then given as
\begin{eqnarray}
E^{(1)}_p(\mathbf x)&=&\sum_n\mathcal E^{(1)}_n f_n(\mathbf x)\simeq-G \sum_n\mathcal E^0_n f_n(\mathbf x) \nonumber\\
&=&-G\mathcal E^0_p f_p(\mathbf x)\equiv \mathcal E^{(1)}_p f_p(\mathbf x)\;.
\end{eqnarray}
From Eqs.\,\eqref{eq10} and \eqref{eq11} we have 
\begin{equation}
\frac{\bar{g}^2}{\gamma_p}=\sum_n\frac{g_n^2}{\gamma_n}=\gamma_r G\;. \label{eq21}
\end{equation}
Therefore,
\begin{displaymath}
\mathcal E^{(1)}_p=-G\mathcal E^0_p=-\left(\frac{\bar{g}^2}{\gamma_p\gamma_r}\right)\left(\frac{\gamma_r}{\bar g}\mathcal{E}_r^{(1)}\right)=-\frac{\bar g}{\gamma_p}\mathcal{E}_r^{(1)}
\end{displaymath}
indicating that the first round correction $\mathcal{E}_p^{(1)}$ in the pump mode can be regarded as a result of tunneling from the regular mode of $\mathcal{E}_r^{(1)}$ with the same coupling constant $\bar g$ and a decay rate $\gamma_p$ associated with the destination mode or the pump mode.

Our identification of $\bar g$ and $\gamma_p$ is not restricted to the first round corrections. By using the physical meaning of $\bar g$ and $\gamma_p$, the second round correction to the regular mode would be given by
\begin{displaymath}
\mathcal{E}_r^{(2)}=\frac{\bar g}{\gamma_r}\mathcal{E}_p^{(1)},
\end{displaymath}
which can be simplified as
\begin{displaymath}
\mathcal{E}_r^{(2)}=\frac{\bar g}{\gamma_r}\left(-\frac{\bar g}{\gamma_p}\right)\mathcal{E}_r^{(1)}=-G\mathcal{E}_r^{(1)},
\end{displaymath}
which is consistent with Eq.\,\eqref{eq-Er1'} obtained without relying on the present identification of $\bar g$ and $\gamma_p$.
Likewise, we expect the second round correction to the pump mode be given by
\begin{displaymath}
\mathcal{E}_p^{(2)}=-\frac{\bar g}{\gamma_p}\mathcal{E}_r^{(2)},
\end{displaymath}
which can be simplified as
\begin{displaymath}
\mathcal{E}_p^{(2)}=-\frac{\bar g^2}{\gamma_r \gamma_p} \mathcal{E}_p^{(1)}=-G\mathcal{E}_p^{(1)}.
\end{displaymath}
This consideration can be extended to all higher-round correction terms. Summing up all correction terms, we obtain
\begin{displaymath}
\mathcal{E}_p=\frac{\mathcal{E}_p^0}{1+G}\;,
\end{displaymath}
which is the same as Eq.\,\eqref{Ep0}, also obtained without using the present identification of $\bar g$ and $\gamma_p$. 
Therefore, we conclude that $\bar g$ is the coupling constant between the pump mode and the regular mode and $\gamma_p$ is the decay rate of the pump mode.

The physical meaning of $G$ and $\gamma'_r$ can be obtained as follows.
According to the cavity quantum electrodynamics (QED), when a high-Q oscillator with a decay rate $\gamma_H$ is coupled with a coupling constant $g$ to a low-Q oscillator with a decay rate $\gamma_L$ ($\gg \gamma_H$), total decay rate of the high-Q mode is enhanced as \cite{c-QED}
\begin{equation}
\gamma'_H=\gamma_H+\frac{g^2}{\gamma_L}=\gamma_H\left(1+\frac{g^2}{\gamma_H \gamma_L}\right)\;. \label{eq22}
\end{equation}
This phenomenon is known as the enhanced spontaneous emission in the cavity QED. Comparing Eqs.\,\eqref{eq22} and \eqref{eq21}, we can identify that the righthand side of Eq.\,\eqref{eq21} is nothing but the increment in the decay rate of the regular mode induced by tunneling into all chaotic modes and that 
$G$ is just an enhancement factor with respect to the uncoupled decay rate $\gamma_r$. Total decay rate of the regular mode modified by the tunneling is then given by $\gamma_r(1+G)$, which was defined as $\gamma'_r$ in Eq.\,\eqref{eq17}. In fact, this total decay rate is correctly reflected in the lineshape $\mathcal L(\delta)$ of the resonant pumping efficiency in Eq.\,\eqref{eq20}. 

\begin{figure}
\includegraphics[width=3.4in]{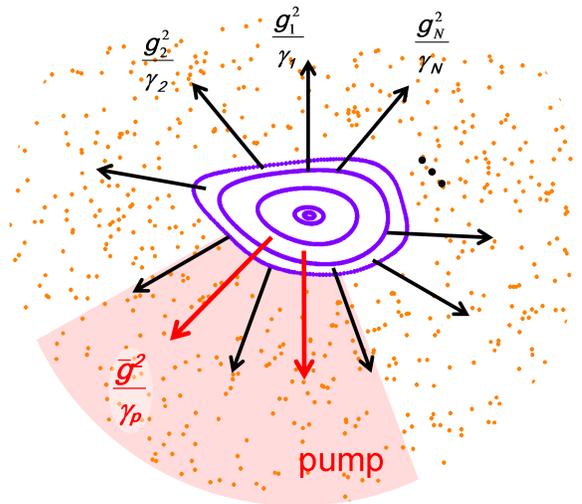}
\caption{The regular mode can couple to all chaotic modes ($n=1,2,\cdots,N$) with an individual tunneling rate $g_n^2/\gamma_n$. We can view this by introducing a pump mode, a specific collection of chaotic modes excited by the pump laser, which has an effective coupling $\bar{g}$ with the regular mode and an effective decay rate $\gamma_p$.}
\label{fig2}
\end{figure}

Equation (\ref{eq21}) shows that the total tunneling rate of the regular mode into all chaotic modes is equal to $\bar g^2/\gamma_p$, which is an effective tunneling rate of the regular mode into the pump mode. This identification is in fact consistent with the Fermi golden rule \cite{Backer08}:
\begin{displaymath}
W=\frac{2\pi}{\hbar}\left|\langle p |\mathcal V|r\rangle \right|^2 \delta_E\;,
\end{displaymath}
where $r$ and $p$ stand for the regular and the pump modes, respectively, $\mathcal V$ is the interaction Hamiltonian, and $\delta_E$ is the density of states.
If $\mathcal V$ is known, we can calculate its matrix element. Let us denote it as
\begin{displaymath}
\langle p |\mathcal V|r\rangle \equiv \hbar \bar g\;.
\end{displaymath}
The density of states is obtained by counting the number of modes per unit energy interval or by taking the inverse of the energy interval associated with the pump mode
\begin{displaymath}
\delta_E=\frac{1}{\hbar}\left[\int_0^\infty \frac{\gamma_p^2}{(\omega-\omega_r)^2+\gamma_p^2}d\omega\right]^{-1}\frac{1}{2}
=\frac{1}{2\pi\hbar\gamma_p}\;.
\end{displaymath}
The extra 1/2 factor comes from the fact that only one polarization direction is allowed for the tunneling process out of two possible polarization directions. Therefore, the tunneling rate into the pump mode given by the Fermi golden rule is
\begin{displaymath}
W=\frac{2\pi}{\hbar} |\hbar\bar g|^2 \frac{1}{2\pi\hbar\gamma_p}=\frac{|\bar g|^2}{\gamma_p}\;.
\end{displaymath}

Lastly, with the above identification of $\gamma_p$ and $\gamma'_r$, the meaning of the factor $\gamma_p/\gamma'_r$ appearing in the pumping efficiency formula, Eq.\,\eqref{eq16}, becomes clear. It is an intensity build-up factor. In non-resonant pumping, the intensity of the pump mode is inversely proportional to its loss rate $\gamma_p$, whereas in the resonant pumping the intensity of the regular mode, as a true eigenmode of the system, is inversely proportional to its total decay rate $\gamma'_r$. So, the factor $\gamma_p/\gamma'_r$ measures how much intensity build-up is enhanced by the resonant pumping compared to the nonresonant pumping.\\ \\

\subsection{Validity of rate equation model}
One may wonder whether the pumping efficiency formula of Eq.\,\eqref{eq16} might also be obtained from rate equations for intensities not electric fields. In order to address this question, let us consider a two-mode problem for simplicity, in which the intensities $I_r$ and $I_c$ in the regular and chaotic modes, respectively, satisfy the following rate equations.
\begin{eqnarray}
\dot{I_c}&=&R-2\gamma_c I_c-\left[\frac{2g^2}{\gamma_r} I_c -\frac{2g^2}{\gamma_c} I_r \right], \label{eq31}\\
\dot{I_r}&=&-2\gamma_r I_r+\left[\frac{2g^2}{\gamma_r} I_c -\frac{2g^2}{\gamma_c} I_r \right], \label{eq32}
\end{eqnarray}
where $R$ is a pumping rate for the chaotic mode. The first term in the square bracket is the energy transfer from the chaotic mode to the regular mode while the second term is that from the regular mode to the chaotic mode. Since $G=g^2/(\gamma_c\gamma_r)$, we can rewrite these equations as
\begin{eqnarray}
\dot{I_c}&=&R-2\gamma_c (1+G) I_c +2\gamma_r G I_r\;, \nonumber \\
\dot{I_r}&=&-2\gamma_r (1+G) I_r +2\gamma_c G I_c\;. \nonumber
\end{eqnarray}
Since rate equations are valid in the limit of large decays, the above equations are valid only when $G\ll 1$.

Nonetheless, the steady-state ($\dot{I_c}=\dot{I_r}=0$) solution is obtained as
\begin{eqnarray}
I_c&=&I_c^0  \frac{1+G}{1+2G}\simeq I_c^0 (1-G), \label{eq33} \\
I_r&=&I_c^0 \frac{\gamma_c}{\gamma_r}\frac{G}{1+2G}\simeq I_c^0 \frac{\gamma_c}{\gamma_r} G, \label{eq34}
\end{eqnarray}
where $I_c^0\equiv R/(2\gamma_c)$. Unlike Eqs.\,\eqref{eq12} and \eqref{eq13}, the results of the mode-mode coupling model, the rate equation model shows that the power removed from the chaotic mode would be all transferred to the regular mode. Moreover, from Eqs.\,\eqref{eq31} and \eqref{eq32} we obtain a statement of energy balance:
\begin{eqnarray}
({\rm pumping})&=&R\nonumber\\
&=&2\gamma_c I_c +2\gamma_r I_r=2\gamma_c I_c^0\nonumber\\
&=&{\rm constant=(energy\;loss)}\;,\nonumber
\end{eqnarray}
which is consistent with Eqs.\ (\ref{eq33}) and (\ref{eq34}).

This energy balance is not satisfied in the mode-mode coupling problem of Eqs.\ (\ref{eq1}) and (\ref{eq2}), which is simplified for two modes of $\mathcal E_c$ and $\mathcal E_r$ on resonance as
\begin{eqnarray}
\dot{\mathcal E}_c&=&a_c E_0 -\gamma_c\mathcal{E}_c -g \mathcal{E}_r\;, \label{eq1'} \nonumber\\
\dot{\mathcal E}_r &=& -\gamma_r\mathcal{E}_r+ g \mathcal{E}_c\;, \label{eq2'} \nonumber
\end{eqnarray}
which can be rewritten in terms of intensities as
\begin{eqnarray}
\dot{I_c}&=&2\Re[a_c E_0\mathcal E_c^*] -2\gamma_c I_c -2g \Re[\mathcal E_r^*\mathcal E_c]\;, \label{eq35}\\
\dot{I_r}&=&-2\gamma_r I_r +2g \Re[\mathcal E_r^*\mathcal E_c]\;, \label{eq36}
\end{eqnarray}
where $\Re[\ldots]$ represents a real component. Comparing these equations with Eqs.\ (\ref{eq31}) and (\ref{eq32}), we identify the following correspondence:
\begin{eqnarray}
2\Re[a_c E_0\mathcal E_c^*] &\leftrightarrow& R\;, \\
2g \Re[\mathcal E_r^*\mathcal E_c]  &\leftrightarrow& \frac{2g^2}{\gamma_r} I_c -\frac{2g^2}{\gamma_c} I_r\;.
\end{eqnarray}
However, the pumping term $2\Re[a_c E_0\mathcal E_c^*]$ is not constant but dependent on $\mathcal E_c$. 
This dependence is caused by the interference of the pump field with the field of the chaotic mode.
Using Eq.\,\eqref{eq4}, we can rewrite the pumping term as
\begin{equation}
2\Re[a_c E_0\mathcal E_c^*]=\frac{2\Re[a_c E_0\mathcal E_c^0]}{1+G}\equiv \frac{R_0}{1+G}\;, \label{eq39}
\end{equation}
which decreases as $G$ increases. In fact, from Eqs.\,\eqref{eq12} and \eqref{eq13} in the steady state, we have
\begin{eqnarray}
({\rm pumping})&=&2\Re[a_cE_0\mathcal E_c^*]\nonumber\\
&=&2\gamma_c I_c +2\gamma_r I_r =\frac{2\gamma_c I_c^0}{1+G} \nonumber
\end{eqnarray}
from which we identify $R_0=2\gamma_c I_c^0$.

The rate equation model and the mode-mode coupling model should give the same result in the limit of large decays or $G\ll 1$. 
However, they give seemingly different results, Eqs.\,\eqref{eq33} and \eqref{eq34} compared to Eqs.\,\eqref{eq12} and \eqref{eq13}. The difference originates from the assumption that the pumping rate would be constant in the rate equation model. We have found in Eq.\,\eqref{eq39} that it varies as $(1+G)^{-1}$. In fact, if we replace the pumping rate $R$ in the rate equation model with $R=2\gamma_c I_c^0/(1+G)$, we then obtain
\begin{eqnarray}
I_c&=&I_c^0\frac{1}{1+2G}\simeq I_c^0(1-2G)\;,\\
I_r&=&I_c^0\frac{\gamma_c}{\gamma_r}\frac{G}{(1+G)(1+2G)}\simeq I_c^0\frac{\gamma_c}{\gamma_r}G\;,
\end{eqnarray}
which is now perfectly consistent with Eqs.\,\eqref{eq12} and \eqref{eq13}, the results of the mode-mode coupling model, in the limit of $G\ll 1$.

\begin{table*}
\begin{tabular}{|c||c|c|c||c|c|c|c|c|c|}
\hline
mode& $\epsilon $ & $\gamma'_r$ & $\beta_r$ & $\gamma_r$ & $G$ & $\bar g$ & $\alpha$&$\alpha'$&$\gamma_r G$\\
 $l$ & {\rm (exp)} & ({\rm exp, in $10^9$ s}$^{-1}$) & {\rm (sim)} & ({\rm in $10^9$ s}$^{-1}$) &  & ({\rm in $10^{10}$ s}$^{-1}$) & & &({\rm in $10^9$ s}$^{-1}$)\\
\hline
1&$25\pm4$&$2.0\pm0.6$&$1.00\pm0.05$&$1.7\pm0.5$& $0.19\pm0.09$ &$2.3\pm0.7$&$0.29\pm0.11$&$0.16\pm0.06$&$0.32\pm0.18$\\
2&$60\pm8$&$7.9\pm2.3$&$0.9\pm0.1$&$4.5\pm1.9$& $0.75\pm0.53$ &$7.6\pm2.9$&$0.67\pm0.20$&$0.43\pm0.17$&$3.4\pm2.6$\\
3&$22\pm4$&$20\pm6$&$0.6\pm0.1$&$15\pm5$& $0.30\pm0.17$ &$8.7\pm2.9$&$0.41\pm0.16$&$0.23\pm0.10$&$4.5\pm2.9$\\
4&$5.5\pm0.7$&$260\pm80$&$0.4\pm0.1$&$190\pm70$&$0.36\pm0.27$ &$34\pm14$&$0.46\pm0.21$&$0.26\pm0.14$&$68\pm57$\\
5&$2.7\pm0.4$&$790\pm230$&$0.3\pm0.1$&$440\pm190$&$0.80\pm0.62$ &$77\pm34$&$0.69\pm0.21$&$0.44\pm0.19$&$350\pm310$\\
\hline
\end{tabular}
\caption{Decay rate $\gamma_r$ of the uncoupled regular mode, the tunneling-induced enhancement factor $G$ in the modified decay rate $\gamma'_r$ of the regular mode, the effective coupling constant $\bar g$ for the pump mode, the coupling efficiencies $\alpha$ and $\alpha'$, and the total tunneling rate $\gamma_r G$ are all obtained from our analysis.}
\end{table*}

\section{Eigenvalue solution} \label{sec4}


The decay rate $\gamma_r$ is the decay rate of the regular  mode when it is uncoupled to other modes. However, the decay rate directly measured in experiments  is not $\gamma_r$, but a modified decay rate \cite{Lee02} due to the tunneling into all chaotic modes. By physical considerations in the preceding section, we identified $\gamma'_r$ defined in Eq.\,\eqref{eq17} is this modified decay rate appearing in the lineshape of the regular mode. 

The expression for the modified decay rate or a total decay rate of the regular mode as a true eigenmode of the system can also be rigorously derived by solving the eigenvalue problem of Eqs.\,\eqref{eq1} and \eqref{eq2} on resonance ($\Delta=0$), which can be rewritten in a matrix form as
\begin{equation}
\frac{d}{dt}{\mathbf E}=-{\mathbf \Gamma}{\mathbf E}+E_0{\mathbf A}\;,
\end{equation}
where
{\setlength\arraycolsep{1pt}
\begin{displaymath}
{\mathbf E}=
\left[ \begin{array}{c}
\mathcal E_r \\
\mathcal E_1 \\
\vdots\\
\mathcal E_N
\end{array} \right],\;\;
{\mathbf \Gamma}=\left[ \begin{array}{cccc}
\gamma_r & -g_1 & \cdots & -g_N \\
g_1 & \gamma_1 & \cdots & 0 \\
\vdots & \vdots & & \vdots\\
g_N & 0 & \cdots & \gamma_N
\end{array} \right],\;\;
{\mathbf A}=\left[ 
\begin{array}{c}
0 \\
a_1 \\
\vdots\\
a_N
\end{array} \right]\;.
\end{displaymath}}
The steady-state solution in Eqs.\,\eqref{eq3} and \eqref{eq4} are obtained by letting the lefthand side vanish and multiplying the inverse matrix ${\mathbf \Gamma}^{-1}$ to both sides: 
\begin{equation}
{\mathbf E}=E_0{\mathbf \Gamma}^{-1}{\mathbf A}\;.
\end{equation}
For transient response of the system we also need to consider the homogeneous equation $\dot{{\mathbf E}}=-{\mathbf \Gamma}{\mathbf E}$. The solution of the homogeneous equation determines the spectrum of the system. For a trial solution, we assume ${\mathbf E}\propto e^{-\lambda t}$, and the result, ${\mathbf \Gamma}{\mathbf E}=\lambda{\mathbf E}$, is nothing but an eigenvalue equation with the eigenvalues obtained from $\det(\mathbf \Gamma-\lambda\mathbf I)=0$. The determinant can be arranged as
\begin{eqnarray}
\lefteqn{\det(\mathbf \Gamma-\lambda\mathbf I)}\nonumber\\
&=&\left[\prod_n^N(\gamma_n-\lambda)\right]\left[(\gamma_r-\lambda)+\sum_n^N\frac{g_n^2}{(\gamma_n-\lambda)}\right].
\end{eqnarray}
So, $\det(\mathbf \Gamma-\lambda\mathbf I)=0$ if $\lambda=\gamma_n$ ($n=1,2,\ldots,N$) or if the quantity in the second bracket vanishes. Since $\gamma_n\gg \lambda\sim \gamma_r$, the expression in the second bracket is simplified as
\begin{equation}
\lambda\simeq \gamma_r+\sum_n^N\frac{g_n^2}{\gamma_n}=\gamma_r(1+G)\;. \label{eq30}
\end{equation}

The eigenvalue $\lambda=\gamma_n$ has the eigenvector $E_j=\delta_{jn}\equiv \psi_n$ (there are $N$ such eigenvectors), which is the original uncoupled $n$th chaotic mode. The eigenvalue $\lambda=\gamma'_r=\gamma_r(1+G)$ corresponds to the modified decay rate of the regular mode. This result is consistent with the one obtained from the cavity-QED and the lineshape considerations in the preceding section. The eigenvector $\psi'_r$ corresponding to this eigenvalue is the modified regular mode and it is easily obtained as
\begin{equation}
\psi'_r=\psi_r-\sum_n \frac{g_n}{\gamma_n}\psi_n
\end{equation}
up to the first order of $g_n/\gamma_n$. Here $\psi_r$ is the uncoupled regular mode. The modified regular mode, which is what we measure in the spectrum as a true eigenmode of the system, has a small contribution from each chaotic mode with a relative amplitude of $(g_n/\gamma_n) \ll 1$.

\section{Analysis} \label{sec5}

In the experiment by Yang {\em et al.}\,\cite{resonant-pumping}, the relative pumping efficiency was measured on resonance, the formulae for which are
\begin{eqnarray}
\epsilon(0)&=&\left[1-\alpha(G)\right]+\frac{\gamma_p\beta_r}{\gamma'_r \beta_p}\alpha'(G),  \quad \quad \mathrm{for}\;\gamma_L \ll \gamma'_r
\nonumber\\
&=&\left[1-\frac{\gamma'_r}{\gamma_L}\alpha(G)\right]+\frac{\gamma_p\beta_r}{\gamma_L\beta_p}\alpha'(G),
\mathrm{for}\;\gamma_L \gg \gamma'_r \nonumber
\end{eqnarray}
The pumping efficiency $\epsilon(0)$ depends on the following parameters: $\gamma_r, \gamma'_r, \gamma_p, \gamma_L, G, \beta_p$ and $\beta_r$, among which $\gamma_p, \beta_p$ and $\beta_r$ are obtained from numerical simulations (see below) and $\gamma_L$ the pump laser linewidth and $\gamma'_r$ the observed linewidth of the regular mode are obtained from experiment. There are left two unknown parameters $\gamma_r$ and $G$. There is one more relation to use, that is, $\gamma'_r=\gamma_r(1+G)$. Therefore, we can simultaneously solve two equations to obtain $\gamma_r$ and $G$. Once they are found, we can then obtain the effective coupling efficiency $\bar g\equiv \gamma_p\sum_n g_n^2/\gamma_n$.

The decay rates of the pump mode $\gamma_p$ can be calculated from the following ray simulation \cite{Yang08}: a bundle of rays is initially prepared in the pumping position with a pumping angle which correspond to the experimental pumping condition and the path length $L_p$ of the bundle of rays before it escapes the cavity is calculated. The decay rate $\gamma_p$ is then simply $c/(2mL_p)$ with $c$ the speed of light and $m$ the refractive index of the cavity medium. 
The overlap factor $\beta_p$ between the pump mode and the lasing mode of interest can also be obtained from a similar ray simulation.
The overlap factor $\beta_r$ between the regular mode and the lasing mode can be obtained from wave calculation, by using polar-angle-averaged mode distributions. Table 1 summarizes the results of the present analysis applied to the experiment of Yang {\em et al.}\,\cite{resonant-pumping}.

\section{Conclusion} \label{sec6}
We have developed a mode-mode coupling theory for the resonant pumping via dynamical tunneling in a deformed microcavity laser. We derived a formula for the pumping efficiency, which is readily measured in experiments, as a function of pump detuning, coupling constants between an uncoupled high-Q regular mode and uncoupled chaotic modes corresponding to the chaotic sea surrounding the regular region supporting the regular mode, and decay rates of the regular and chaotic modes. Analytic expressions for coupling efficiency of the pump into the regular mode is derived. It is shown that the pump-excited chaotic modes collectively couple with the regular mode with an effective coupling constant and a single effective decay rate. We show that the coupling efficiency and the effective coupling constant can be obtained from the observed pumping efficiency on resonance by using our theory. We applied our theory to the experiment by Yang {\em et al.} and obtained the coupling efficiency as well as the effective coupling constant as an example.

\begin{acknowledgments}
This work was supported by NRL and WCU Grants from National Research Foundation of Korea.
\end{acknowledgments}

\end{document}